\documentclass[aps,prd,showpacs,nofootinbib,superscriptaddress,preprint,tightenlines]{revtex4}
\usepackage{graphicx}
\def\be{\nopagebreak[3]\begin{equation}}
\def\ee{\end{equation}}
\def\ba{\nopagebreak[3]\begin{eqnarray}}
\def\ea{\end{eqnarray}}

\newcommand{\teta}{\rlap{\lower2ex\hbox{$\,\tilde{}$}}\eta{}}

\begin{document}
\preprint{\vbox{\baselineskip=12pt \rightline{ICN-UNAM-06/04}
\rightline{gr-qc/0605014} }}
\title{Quantum geometry and microscopic black hole entropy}
\author{Alejandro Corichi}
\email{corichi@matmor.unam.mx} \affiliation{Instituto de
Matem\'aticas, Universidad Nacional Aut\'onoma de M\'exico, A.
Postal 61-3, Morelia, Michoac\'an 58090, Mexico}
\affiliation{Instituto de Ciencias Nucleares,
 Universidad Nacional Aut\'onoma de M\'exico,
A. Postal 70-543, M\'exico D.F. 04510, Mexico}

\author{Jacobo D\'\i az-Polo}
\email{Jacobo.Diaz@uv.es}\affiliation{Departamento de Astronom\'\i
a y Astrof\'\i sica, Universidad de Valencia, Burjassot-46100,
Valencia, Spain}

\author{Enrique Fern\'andez-Borja}\email{Enrique.Fernandez@uv.es}
\affiliation{Departamento de F\'\i sica Te\'orica and IFIC, Centro
Mixto Universidad de Valencia-CSIC. Universidad de Valencia,
Burjassot-46100, Valencia, Spain}

\begin{abstract}
Quantum black holes within the loop quantum gravity (LQG)
framework are considered. The number of microscopic states that
are consistent with a black hole of a given horizon area $A_0$ are
counted and the statistical entropy, as a function of the area, is
obtained for $A_0$ up to $550\, \ell^2_{\rm Pl}$. The results are
consistent with an asymptotic linear relation and a logarithmic
correction with a coefficient equal to $-1/2$. The Barbero-Immirzi
parameter that yields the asymptotic linear relation compatible
with the Bekenstein-Hawking entropy is shown to coincide with a
value close to $\gamma=0.274$, which has been previously obtained
analytically. However, a new and oscillatory functional form for
the entropy is found for small, Planck size, black holes that
calls for a physical interpretation.
\end{abstract}

 \maketitle

\section{Introduction}

 One of the most crucial test that a
candidate quantum theory of gravity must pass is to provide a
mechanism to account for the microscopic degrees of freedom of
black holes (BH). For more than 30 years this has been a
theoretical challenge ever since the discovery by Bekenstein and
Hawking that black holes are quantum in nature \cite{BH}. It is
not unfair to say that at the moment we have only two candidates
for quantum gravity that have offered such an explanation:
string/brane theory \cite{vafa} and loop quantum gravity
\cite{carlo,ABCK}. The LQG formalism allows to include several
matter couplings (including non-minimal couplings) and black holes
far from extremality, in four dimensions.  The approach uses as
starting point isolated horizon (IH) boundary conditions at the
classical level, where the interior of the BH is excluded from the
region under consideration. In this sense, the description is
somewhat effective, since part of the information about the interior is
encoded in the boundary conditions at the horizon that in the quantum
theory get
promoted to a condition that horizon states must satisfy.
There is also an important issue
regarding this formalism. Loop quantum gravity possesses a one
parameter family of inequivalent representations of the classical
theory labelled by a real number $\gamma$, the so called
Barbero-Immirzi (BI) parameter  (it is
the  analogue of the $\theta$ ambiguity in QCD \cite{BI}). It
turns out that the BH entropy calculation provides a linear
relation between entropy and area for very large black holes (in
Planck units) as,
$$S=\lambda\,A(\gamma),$$
 where the parameter
$\lambda$ is independent of $\gamma$ and depends only in the
counting. We have put the $\gamma$ dependence in the Area, since
the parameter appears explicitly in the area spectrum. The
strategy that has been followed within the LQG community is to
regard the Bekenstein-Hawking entropy $S=A/4$ as a requirement
that large black holes  should satisfy. This fixes uniquely the
value of $\gamma=\gamma_0$ once and for all, by looking at the
asymptotic behavior, provided that one has the `correct counting'
that provides the right value for $\gamma_0$. Whenever we have
independent tests that call for a specific value of $\gamma$, one
better find out that the value $\gamma_0$ `works', or else LQG
would be in trouble.\footnote{The presumed relation between
quasinormal modes and quantum black hole transition \cite{olaf},
that calls for a different value of $\gamma$, even when
intriguing, does {\it not} constitute an independent test from our
perspective.}

The parameter $\lambda$ above depends on the calculation of the
entropy, that is, in the counting of states compatible with
whatever requirements we have imposed. The matter has not been
free from some controversy. In the original calculations (once
isolated horizons were understood to be vital) \cite{ABCK,ABK},
the number of states was underestimated; the entropy appeared to
arise from a special set of states where the contribution to the
area from each puncture was the same and corresponded to that of the
minimum spin possible. Later on, it was realized that this
calculation had failed to consider many states \cite{Dom:Lew}, and
a corrected analytical estimation of entropy, and value of the
Barbero-Immirzi parameter, was proposed in \cite{Meiss}.
Furthermore, a still different calculation appeared soon afterward
\cite{GM} which gave yet another value for $\gamma$. This
situation suggests that a clear understanding of the black hole formalism and
entropy counting within LQG is needed. In a recent detailed
analysis of the existing countings, it has been shown that an
unambiguous analytical calculation that yields the `correct' value
of the Immirzi parameter indeed exists (see \cite{CDF} for
details). Of course, one could always remain skeptical and ask for
an `acid test' of the formalism, and the counting.

The simplest such test would be to just count states. The purpose
of this letter is to do that. We count states, by means of a
simple algorithm, of a quantum black hole within the existing
formalism in LQG \cite{ABCK}, compatible with the restrictions
that this framework imposes, as in \cite{ABK}. To be more precise,
we restrict attention to spherical horizons (for which area is the
only free parameter classically) of a fixed horizon area $A_0$ and
compute the number of allowed quantum states, within an interval
$[A_0-\delta A,A_0+\delta A]$ that satisfy the following: i) The
quantum area expectation value satisfies: $\langle\hat{A}\rangle
\in [A_0-\delta A,A_0+\delta A]$, and ii) for which a restriction
on the quantum states of the horizon, $\sum_i m_i=0$, is imposed.
For details see \cite{CDF}. This last `projection constraint'
comes from the consistency conditions for having a quantum horizon
that has, furthermore, the topology of a two-sphere
(it is the quantum analogue of the
Gauss-Bonnet theorem). In the analytical treatments, it has been
shown in detail that, for large black holes (in Planck units), the
entropy behaves as:
$$S=\frac{A}{4}-\frac{1}{2}\ln{A},$$ provided the Barbero-Immirzi
parameter $\gamma$ is chosen to coincide with the value $\gamma_0$,
that has to satisfy \cite{GM}:
\be
 1=\sum_i
(2j_i+1)\,\exp{\left(-2\pi\gamma_0\sqrt{j_i(j_i+1)}\right)}\, .
\ee
The solution to this transcendental equation is approximately
$\gamma_0=0.27398\ldots$ \cite{kripi,GM,CDF}.

Thus, there are two kind of tests one can make. The first one
involves the linear relation between entropy and area that
dominates in the large area regime. This provides a test for the
value of the BI parameter. The second test has to do with the
coefficient of the logarithmic correction (-1/2), a subject that
has had its own share of controversy. The analytical results show
that this coefficient is independent of the linear coefficient and
arises in the counting whenever the $\sum m=0$ constraint is
imposed.

In order to test the validity of the logarithmic correction and
its relation to the constraint, we fix the value of the parameter
$\gamma=\gamma_0$ and compute the number of states, both with and
without the projection constraint. We subtract this functions and
compare the difference with logarithmic functions. We look for
the coefficient that provides the best fit. Once the logarithmic
coefficient is found and the independence of the asymptotic linear
coefficient is established, we perform a variety of countings for
different values of $\gamma$, both with and without the projection
constraint, and consider the slope $c$ of the resulting relation
$S=c\,A$, as a function of $\gamma$. For the function $c(\gamma)$
we look for the value of $\tilde{\gamma}$ for which
$c(\tilde{\gamma})=1/4$.

Another separate issue that one would like to consider is the
applicability of the formalism for `small' black holes. The
isolated horizons boundary conditions are imposed classically in
the variational principle, which means that the horizon is assumed
to exist as a classical object. A natural question is whether the
resulting picture can be trusted for small black holes, not far
from the Planck regime where strong quantum gravity effects can be
expected to appear. Another related question one might try to
answer is the `scale' at which the quantum horizon entropy
approaches the expected form derived from semi-classical/large
horizon area approximation. As we shall see, even when the limited
computing power at our disposal, we shall be able to partially
answer some of these questions.

The remainder of this note is as follows. In Sec.~\ref{sec:2} we
shall describe the algorithm that implements the counting of
states. Sec.~\ref{sec:3} is devoted to describing the results
found. We end this letter with a discussion in Sec.~\ref{sec:4}.

\section{The Counting}
\label{sec:2}

Counting configurations for large values
of the area (or mass) is extremely difficult for the simple reason
that the number of states scales exponentially. Thus, for the
 computing power at our disposal,
we have been able to compute states up to a value of area of about
$A=550\; l^2_P$ (recall that the minimum area gap for a spin $1/2$
edge is about $a_0\approx 6\,l^2_P$, so the number of punctures on
the horizon is  below 100).
At this point the number of
states exceeds $2.8\times 10^{58}$. In terms of Planck masses, the
largest value we have calculated is $M=3.3 \,M_P$. When the projection
constraint is introduced, the upper mass we can calculate is much
smaller, given the computational complexity of implementing the
condition. In this case, the maximum mass is about $1.7\,M_P$.

It is important to describe briefly what the program for counting
does. What we are using is what it is known, within combinatorial
problems, as a brute force algorithm. This is, we are simply
asking the computer to perform all possible combinations of the
labels we need to consider, attending to the distinguishability
-indistinguishability criteria that are relevant \cite{ABK,CDF},
and to select (count) only those that satisfy the conditions
needed to be considered as permissible combinations, i.e., the
area condition and the spin projection constraint. An algorithm of
this kind has an important disadvantage: it is obviously not the
most optimized way of counting and the running time increases
rapidly as we go to little higher areas. This is currently the
main limitation of our algorithm. But, on the other hand, this
algorithm presents a very important advantage, and this is the
reason why we are using it: its explicit counting guarantees us
that, if the labels considered are correct, the result must be the
right one, as no additional assumption or approximation is being
made. It is also important to have a clear understanding that the
algorithm does not rely on any particular analytical counting
available. That is, the program counts states as specified in the
original ABK formalism \cite{ABK}. The computer program has three
inputs: i) the classical mass $M$ (or area $A_0=16\pi\,M^2$), ii)
The value of $\gamma$ and iii) The size of the interval $\delta
A$.

\begin{figure}
  \includegraphics[angle=270,scale=.40]{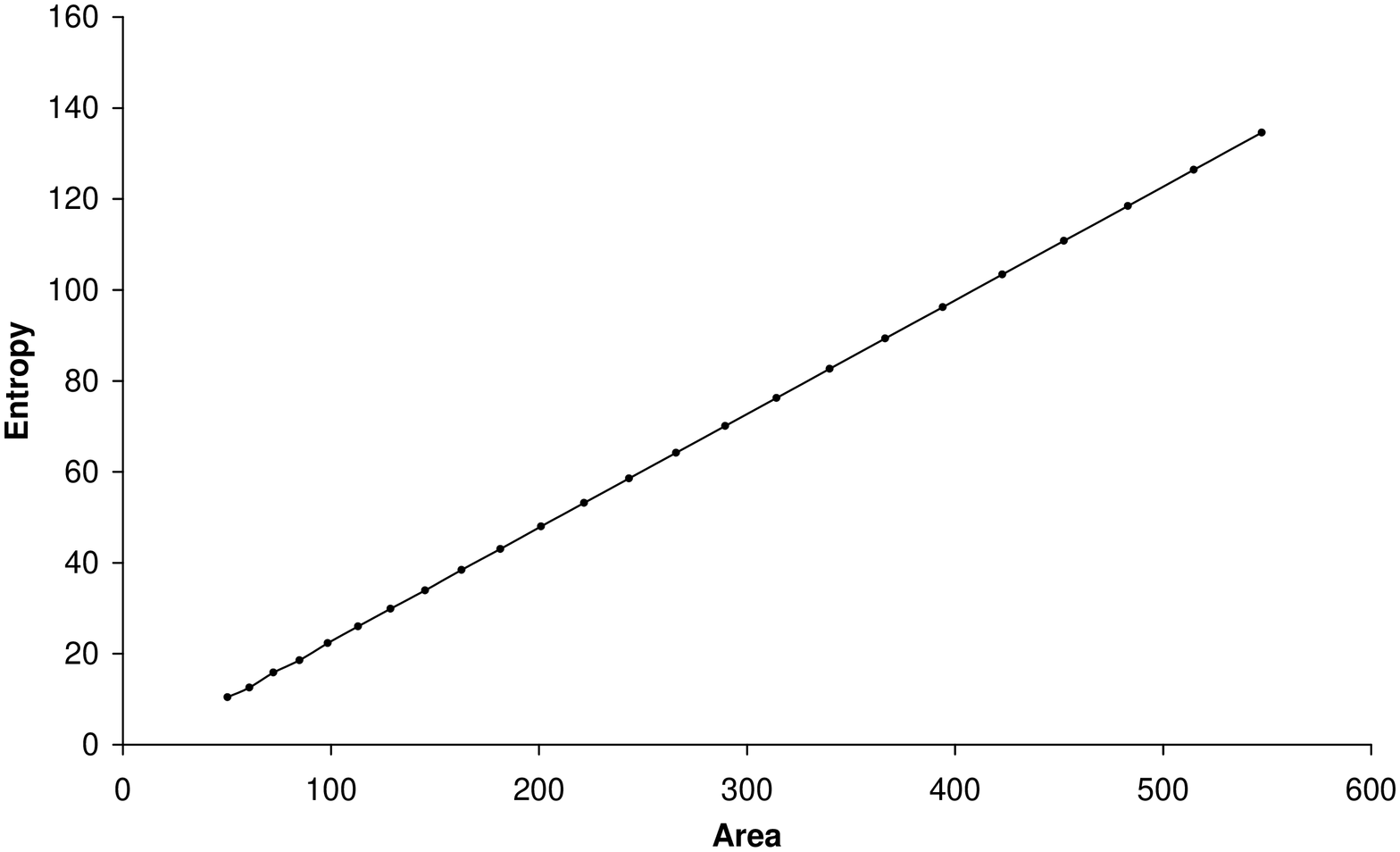}
\caption{\label{fig:1} The entropy as a function of area is shown,
where the projection constraint has not been imposed. The BI is
taken as $\gamma=0.274$.}
\end{figure}

Once these values are given, the algorithm computes the level of
the horizon Chern Simons theory $k=\left[A_0/4\pi\gamma\right]$
and the maximum number of punctures possible $n_{\rm
max}=\left[A_0/4\pi\gamma\sqrt{3}\right]$ (where $[\cdot]$,
indicates the principal integer value). At first sight we see
that the two conditions we have to impose to permissible
combinations act on different labels. The area condition acts over
$j$'s and the spin projection constraint over $m$'s. This allows
us to first perform combinations of $j$'s and select those
satisfying the area condition. After that, we can perform
combinations of $m$'s only for those combinations of $j$'s with
the correct area, avoiding some unnecessary work. We could also be
allowed to perform the counting without imposing the spin
projection constraint, by simply counting combinations of $j$'s
and including a multiplicity factor of $\prod_i (2 j_i + 1)$ for
each one, accounting for all the possible combinations of $m$'s
compatible with each combination of $j$'s. This would reduce
considerably the running time of the program, as no counting over
$m$'s has to be done, and will allow us to separate the effects of
the spin projection constraint (that, as we will see, is the
responsible of a logarithmic correction). It is very important to
notice at this point that this separation of the counting is
completely compatible with the distinguishability criteria.

The next step of the algorithm  is to consider, in increasing
order, all the possible number of punctures (from 1 to $n_{\rm
max}$) and in each case it considers all possible values of $j_i$.
Given a configuration $(j_1,j_2,\ldots,j_n)$ ($n\leq n_{\rm
max}$), we ask whether the quantum area eigenvalue $A=\sum_i
8\pi\gamma\sqrt{{j_i}\left({j_i}+1\right)}$ lies within
$[A_0-\delta A,A_0+\delta A]$. If it is not, then we go to the
next configuration. If the answer is positive, then the labels
$m$'s are considered as described before. That is, for each of
them it is checked whether $\sum m_i=0$ is satisfied.

\begin{figure}
  \includegraphics[angle=270,scale=.40]{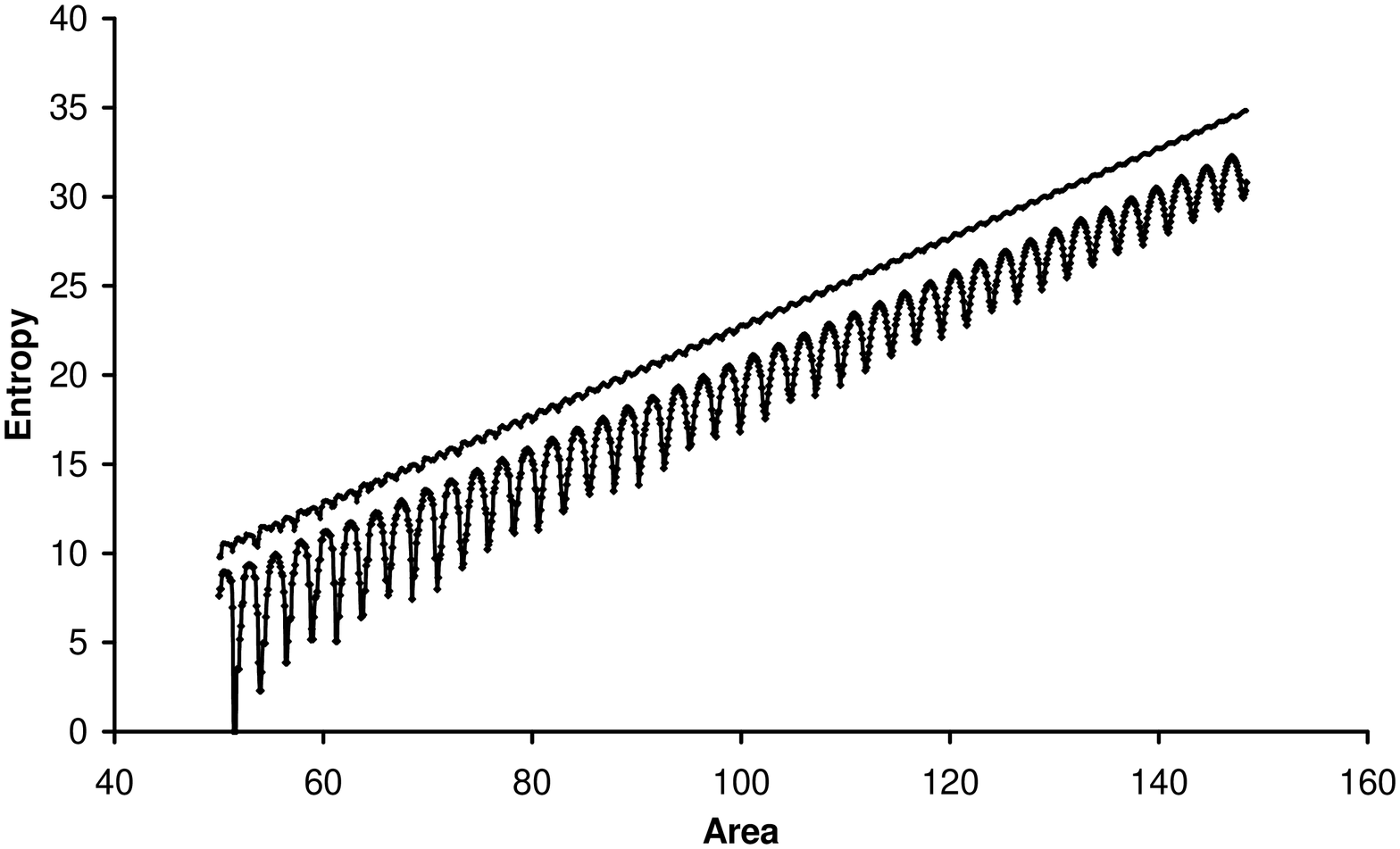}
\caption{\label{fig:2} Entropy {\it vs} Area with and without the
projection constraint, with $\delta A=0.5$.}
\end{figure}

\section{Results}
\label{sec:3}

Let us now present the results found. We shall separate this
section in two parts. In the first one, we shall focus on the
Logarithmic correction, that is, in the results obtained when
considering the spin constraint. In the second part, we shall
report on the asymptotic linear relation that yields information
about the Barbero-Immirzi parameter.

\subsection{Logarithmic Correction}


In Figure~\ref{fig:1}, we have plotted the entropy, as
$S=\ln(\#\,{\rm states})$ {\it vs} the area $A_0$, where we have
counted all possible states without imposing the $\sum m_i$
constraint, and have chosen a $\delta A=0.5$. As it can be seen,
the relation is amazingly linear, even for such small values of
the area. When we fix the BI parameter to be
$\gamma=\gamma_0=0.274$, and approximate the curve by a linear
function, we  find that the best fit is for a slope equal to
$0.2502$.

\begin{figure}
  \includegraphics[angle=270,scale=.40]{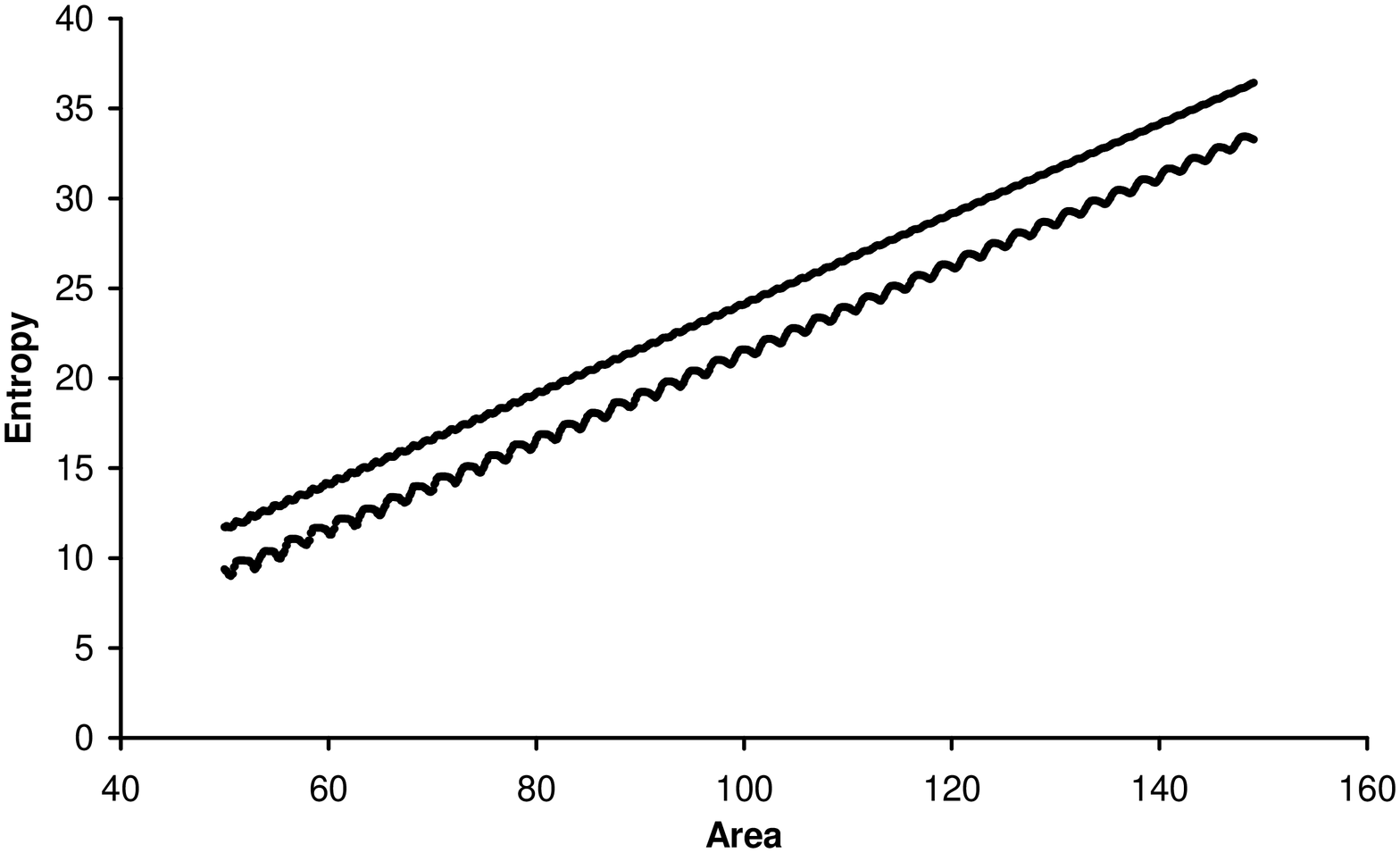}
\caption{\label{fig:3} Entropy {\it vs} Area with and without the
projection constraint, with $\delta A=2$.}
\end{figure}
\begin{figure}
  \includegraphics[angle=270,scale=.40]{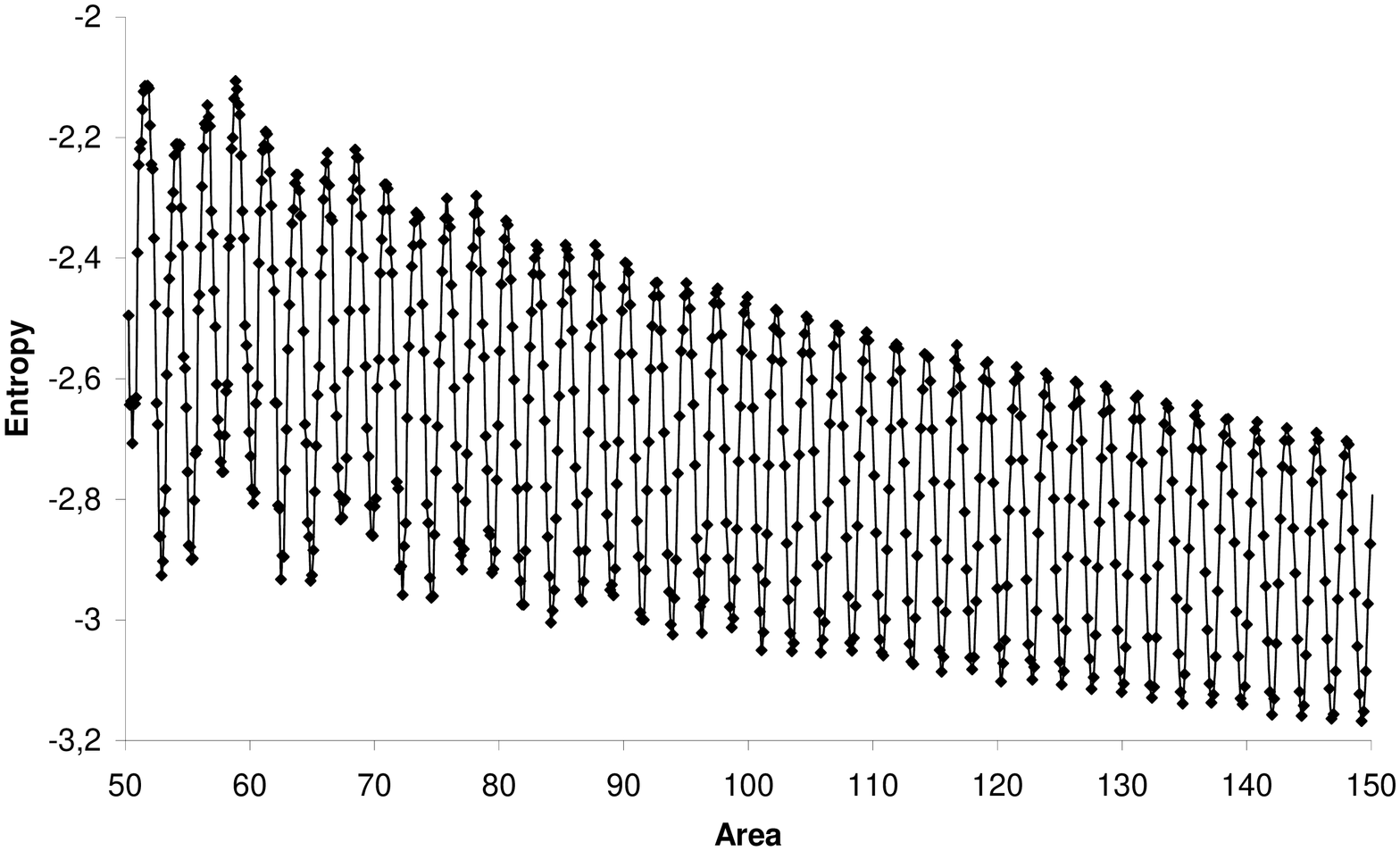}
  \includegraphics[angle=270,scale=.40]{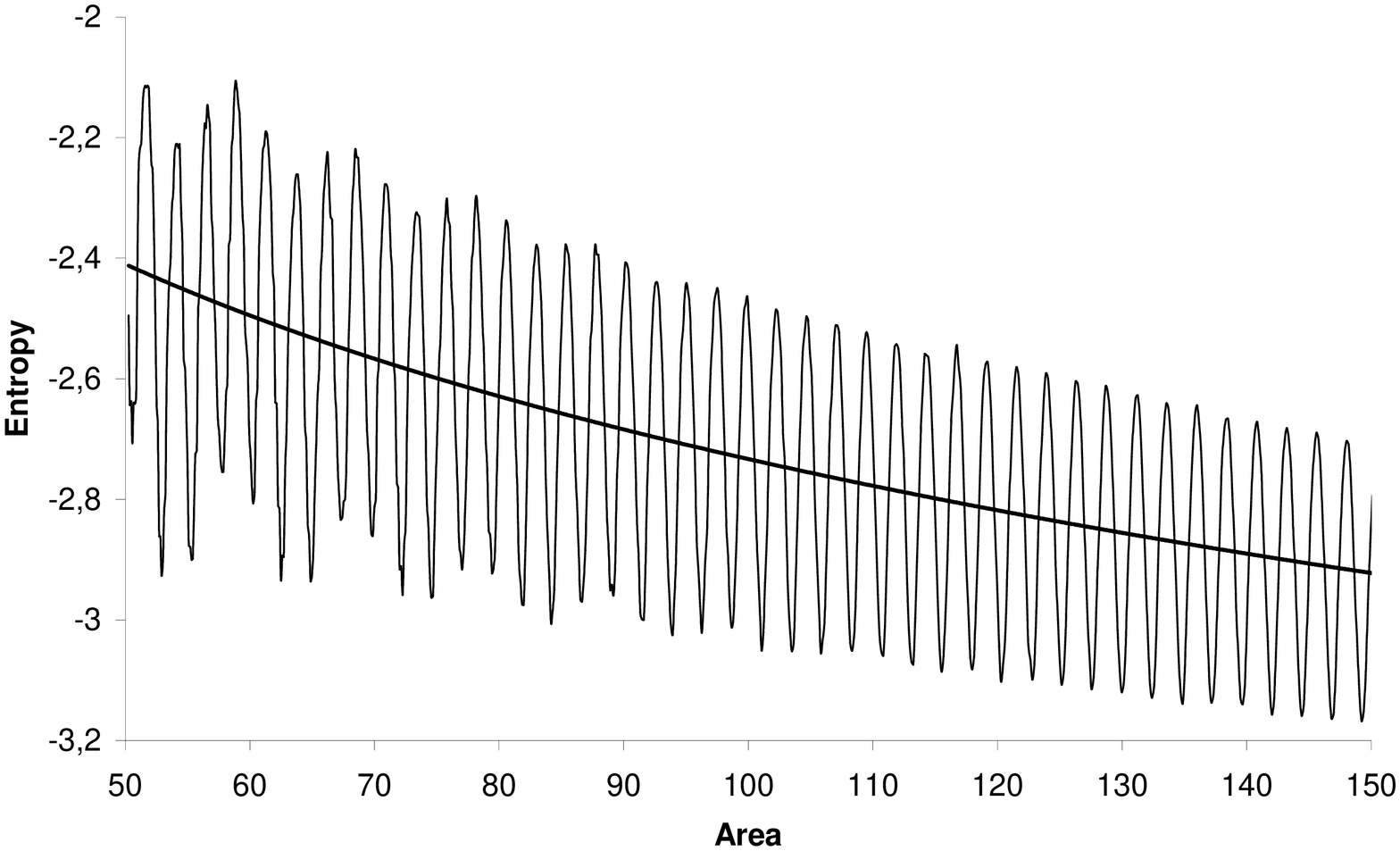} \caption{\label{fig:5}
  The curves of Fig.~\ref{fig:3}
are subtracted and the difference, an oscillatory function, shown
in the upper figure. The curve is approximated by a logarithm
curve in the lower figure.}
\end{figure}

When we include the projection constraint, the computation becomes
more involved and we are forced to consider a smaller range of
values for the area of the black holes. In Figure~\ref{fig:2}, we
plot both the entropy without the projection and with the
projection, keeping the same $\delta A$. The first thing to note
is that for the computation with the constraint implemented, there
are some large oscillations in the number of states. Fitting a
straight line gives a slope of $0.237$. In order to reduce the
oscillations, we increased the size of $\delta A$ to $\delta A=2$.
The result is plotted in Figure~\ref{fig:3}. As can be seen the
oscillations are much smaller, and the result of implementing the
constraint is to shift the curve down (the slope is now 0.241). In
order to compare it with the expected behavior of
$S=A/4-(1/2)\ln{A}$, we subtracted both curves of
Figure~\ref{fig:3}, in the range $A=[50,160]$, and compared the
difference with a logarithmic function. The coefficient that gave
the best fit is equal to $-0.4831$ (See Figure~\ref{fig:5}). What
can we conclude from this? While it is true that the rapidly
oscillating function is far from the analytic curve, it is quite
interesting that the oscillatory function follows a logarithmic
curve with the ``right" coefficient. It is still a challenge to
understand the meaning of the oscillatory phase. Even when not
conclusive by any means, we can say that the counting of states is
consistent with a (n asymptotic) logarithmic correction with a
coefficient equal to (-1/2).

\subsection{Barbero-Immirzi parameter}

\begin{figure}
  \includegraphics[angle=270,scale=.40]{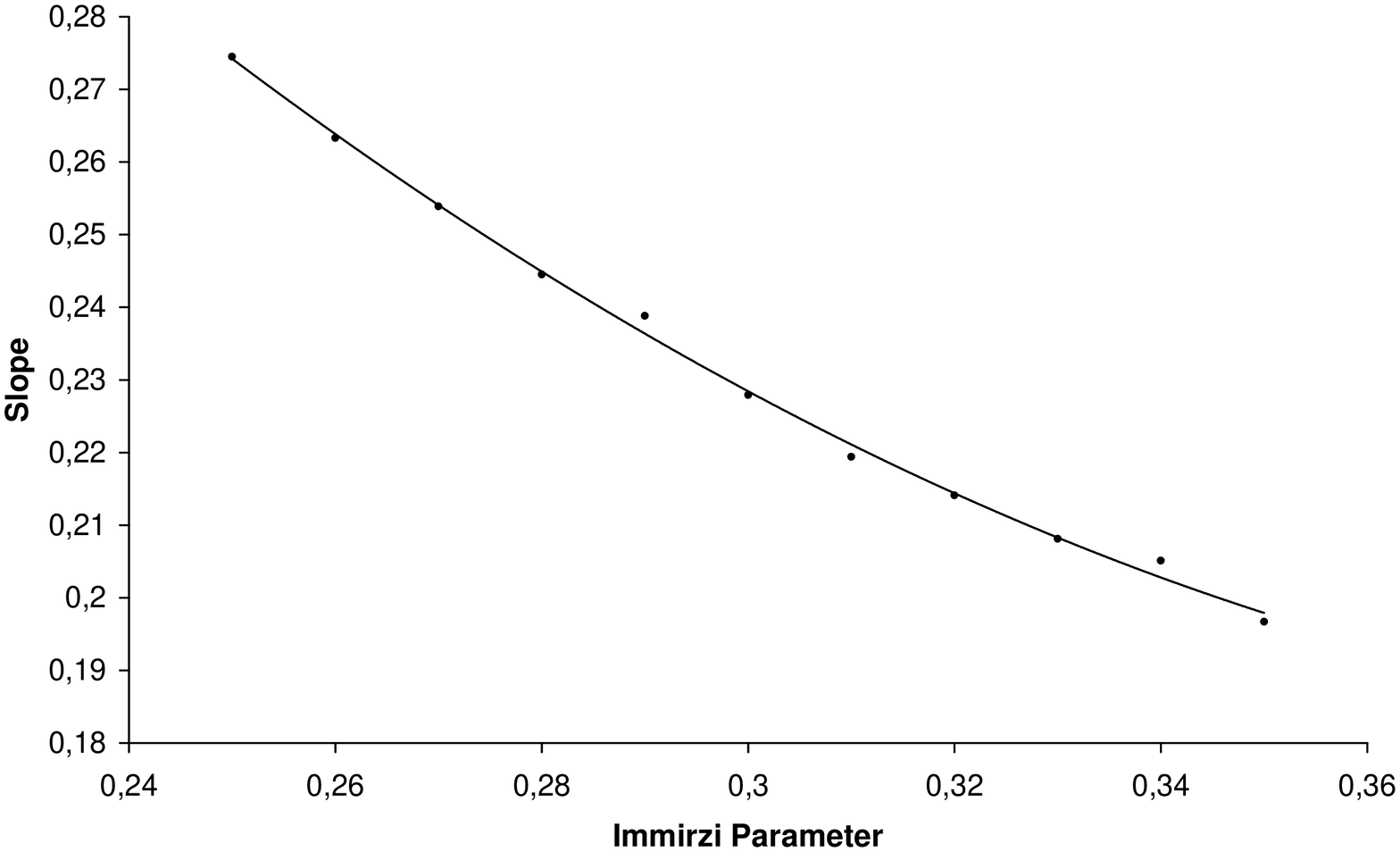}
\caption{\label{fig:6} The slope of the entropy area line is
plotted as function of the Immirzi parameter without the
projection constraint.}
\end{figure}

Let us now assume that
the logarithmic correction is indeed there and that, as the
analytical calculations suggest \cite{GM,CDF}, the projection
constraint does not have any affect on the coefficient of the
linear term, that is, on the Barbero-Immirzi parameter. With this
in mind, we have performed a variety of countings for different
values of $\gamma$, without the projection constraint, and
considered the slope $c$ of the resulting relation $S=c\,A$, as a
function of $\gamma$. For the resulting function $c(\lambda)$ we
looked for the value of $\tilde{\gamma}$ for which
$c(\tilde{\gamma})=1/4$.
This is shown in Figure~\ref{fig:6}.

In order to find this value, we have interpolated the curve and
found the value $\tilde{\gamma}=0.2743691$ for which the slope
is equal to 1/4. It is hard not to note that the value of
$\tilde{\gamma}$ is amazingly close to the value $\gamma_0$ found
analytically.

\begin{figure}
  \includegraphics[angle=270,scale=.40]{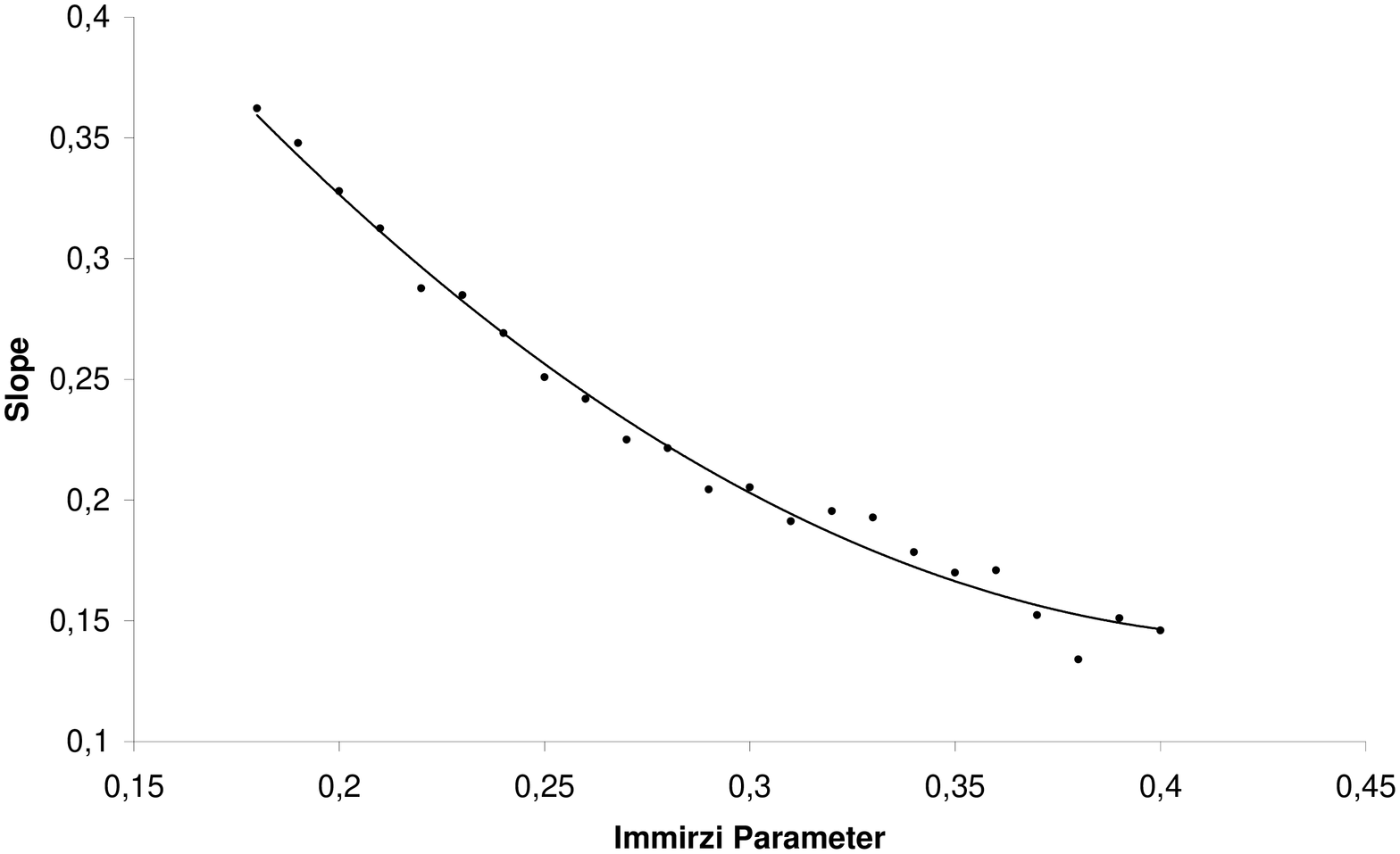}
\caption{\label{fig:7} The slope of the entropy area curve is
plotted as function of the Immirzi parameter with the projection
constraint implemented.}
\end{figure}

When we repeat this procedure  including the $\sum m$ constraint,
just to have a rough idea, we have computed for a limited range of
mass (in steps of 0.1) and for a variety of $\gamma$ in
$[0.18,0.4]$, in steps of $0.01$ and have plotted the results in
Figure~\ref{fig:7}. The value $\gamma'$ where the curve crosses
$1/4$ is $\gamma'=0.2552$, which is still far from the GM value
(which one expects to get for larger BH), but is clearly very far
from the value given in \cite{Meiss}. What is amazing is that,
even when considering only these Plank size horizons, one can
confidently say that there is an asymptotic linear relation
between entropy and area and that the relevant coefficient is
consistent {\it only} with the value of the BI parameter
$\gamma_0=0.27398\ldots$, as found in \cite{GM,CDF}.

\section{Discussion and Outlook}
\label{sec:4}


Within loop quantum gravity, the issue of which states
should be counted when computing the black hole entropy is a
pressing one. The formalism for treating boundary conditions and
the quantum horizon geometry established in \cite{ABCK,ABK}
provides a clear and precise framework. This includes an
unambiguous answer as to which states are to be distinguished and
which are to be considered undistinguishable, and to the question
of which quantum numbers ($j_i$ and/or $m_j$) are to be considered.
In this letter, we have followed a
direct application of the formalism and have counted,
using a simple algorithm, the states that satisfy the conditions
and that yield an area close to a specified value $A_0$.
What we have learned can be summarized as follows:

i) When we do not impose the projection constraint, we find that
very rapidly, the entropy area relation becomes linear.

ii) The BI parameter that yields the desired agreement with
$S=A/4$ is given by the value $\gamma_0=0.27398 \ldots$, and not
by any of the other values found in the literature.

iii) When the projection constraint is incorporated, which analytically
gives the logarithmic correction, the curve gets shifted down and exhibits
some oscillations, but follows on average the expected curve with the
predicted coefficient $-1/2$.

iv) For the rather small value of the BH area computed, and for
$\gamma=\gamma_0$, the total entropy seems to approach a linear
relation with a ratio $S/A$ approaching $1/4$ from below, which is
what one expects due to the logarithmic correction.


It is important to emphasize that the procedure followed here, in
the algorithm implemented, is \emph{not} assuming any of the
analytical estimations available, but rather performing a direct
counting by \emph{brute force} of the microstates consistent with
the macroscopic requirements and thus, responsible for the Black
Hole entropy. In a sense, the results here presented can be seen
as providing strong evidence for which the correct analytical
counting is.

Even when the results presented here shed light on the relation
between  entropy and area, and the Barbero-Immirzi parameter, one
still needs more work to have completely conclusive results. In
particular, one needs more computing capacity to go further in the
range of values analyzed.

Furthermore, the oscillations found in the entropy area relation
certainly call for an explanation. For instance, it is important
to determine whether there there is some area scale set by the
oscillations found in the entropy area relation. To this effect,
we have found the frequency that best approximates the
oscillations, and the frequency in areas gives an area scale of
$\delta A_{\rm osc} = 2.407\,\; \ell^2_{\rm Pl}$. It remains a
challenge to find an explanation for this scale.

It could also happen, for instance, that the thermodynamic
quantities such as temperature (that is usually associated with
$T=\partial M/\partial S$), and the specific heat get modified as
one approaches the Planck scale. The usual, classical relation
between mass and entropy (using the relation S=A/4) implies that a
Schwarzschild black hole has negative specific heat; as the energy
of the Black Hole decreases, the temperature increases, making the
system unstable. One could imagine, for instance, that the
oscillations here found (that are seen to decrease for larger
black holes), make the specific heat positive as one decreases the
area for some (small) value and, thus, would `stabilize' the black
hole. Another intriguing possibility would be to learn something
from this formalism (tailored for large equilibrium systems),
about the evaporation process of small black holes and the issue
of information loss. We shall leave these issues for future
publications.

\section*{Acknowledgments}

\noindent We thank J. Olivert for discussions.
This work was in part supported by DGAPA-UNAM IN108103, CONACyT
U47857-F, ESP2005-07714-C03-01 and FIS2005-02761 (MEC) grants. J.D. thanks MEC for a FPU fellowship.

\end{document}